\documentclass[runningheads]{llncs}
\usepackage{calc}
\usepackage{tikz}
\usetikzlibrary{decorations.markings}
\tikzstyle{vertex}=[circle, draw, inner sep=0pt, minimum size=6pt]
\newcommand{\vertex}{\node[vertex]}
\usepackage{graphicx}
\usepackage{amsmath}
\usepackage{amsfonts}
\usepackage{xcolor}
\usepackage{amssymb}
\usepackage{graphicx}
\usepackage{tikz}
\usetikzlibrary{circuits.logic.US}
\usetikzlibrary{positioning}
\usetikzlibrary{matrix}
\usepackage{graphicx}
\usepackage{amsmath}
\newcommand{\boundellipse}[3]% center, xdim, ydim
{(#1) ellipse (#2 and #3)
}
\definecolor{magenta}{rgb}{0.8, 0.0, 0.8}
\definecolor{cyan}{rgb}{0.0, 1.0, 1.0}
\definecolor{green1}{rgb}{0.1, 0.6, 0.01}
\definecolor{green}{rgb}{0.11, 0.35, 0.02}
\definecolor{brown}{rgb}{0.65, 0.16, 0.16}

\begin{document}

\title{The Satisfactory Partition Problem  }
\author{Ajinkya Gaikwad\inst{1} \and Soumen Maity\inst{1}  \and Shuvam Kant Tripathi\inst{1}}
\authorrunning{A.\,Gaikwad et al.}
\institute{Indian Institute of Science Education and Research\\ Pune, 411008, India \\
%\and
%Nagoya University, Furocho, Chikusa-ku, Nagoya, 464-8601, Japan\\
%\authorrunning{ S.\,Maity, A.\,Gaikwad, and S.K.\,Tripathi}
\email{\texttt{soumen@iiserpune.ac.in}}
\email{\texttt{\{ajinkya.gaikwad|tripathi.shuvamkant\}@students.iiserpune.ac.in}}
}
\maketitle              % typeset the header of the contribution
\begin{abstract} The {\sc Satisfactory Partition} problem consists in deciding if
the set  of vertices of a given undirected graph can be partitioned
into  two nonempty parts  such that each vertex has at least as many neighbours in its part as in
the other part.  This problem was introduced by Gerber and Kobler [European J. Oper. Res. 125 (2000) 
283-291] and further studied by other authors, but its parameterized complexity remains open 
until now. 
It is known that the {\sc Satisfactory Partition} problem, as well as a variant where the parts are required to be of the
same cardinality, are NP-complete.  We enhance our understanding of the problem from the 
viewpoint of parameterized complexity by showing that (1) the problem is FPT when parameterized by 
the neighbourhood diversity of the input graph, (2) it can be solved in $O(n^{8 {\tt cw}})$ where ${\tt cw}$ is the clique-width, 
 (3) a generalized version of the problem is W[1]-hard when parameterized by the treewidth.

\end{abstract}

\section{Introduction}

Gerber and Kobler \cite{GERBER2000283} introduced the problem of deciding if a given graph has a vertex 
partition  into two non-empty parts such that each vertex has at least as many neighbours 
in its part as in the other part. A graph satisfying this property is 
called {\it partitionable}. For example, complete graphs, star graphs, complete bipartite graphs with 
at least one part having odd size are not partitionable, where as some graphs are easily
partitionable: cycles of length at least 4, trees that are  not  star graphs \cite{BAZGAN20061236}. 
\par Given a graph $G=(V,E)$ and a subset $S\subseteq V(G)$, we denote by $d_S(v)$ the degree
of a vertex $v\in V$ in $G[S]$, the subgraph of $G$ induced by $S$. For $S=V$, the subscript is
omitted, hence $d(v)$ stands for the degree of $v$ in $G$. In this paper, we study the parameterized
complexity of {\sc Satisfactory Partition} and {\sc Balanced Satisfactory Partition} problems. We define these 
problems as follows:
 \vspace{3mm}
    \\
    \fbox
    {\begin{minipage}{33.7em}\label{SP1}
       {\sc  Satisfactory Partition}\\
        \noindent{\bf Input:} A graph $G=(V,E)$.\\
    \noindent{\bf Question:} Is there a nontrivial partition $(V_1,V_2)$ of $V$
    such that for every $v\in V$, if $v\in V_i$ then $d_{V_i}(v)\geq d_{V_{3-i}}(v)$?
    \end{minipage} }\\
    
\noindent A variant of this problem where the two parts have equal size is:
\vspace{3mm}
    \\
    \fbox
    {\begin{minipage}{33.7em}\label{BSP}
       {\sc Balanced  Satisfactory Partition}\\
        \noindent{\bf Input:} A graph $G=(V,E)$ on an even number of vertices.\\
    \noindent{\bf Question:} Is there a nontrivial partition $(V_1,V_2)$ of $V$
    such that $|V_1|=|V_2|$ and  for every $v\in V$, if $v\in V_i$ then $d_{V_i}(v)\geq d_{V_{3-i}}(v)$?
    \end{minipage} }\\

Given a partition $(V_1,V_2)$, we say that a vertex $v\in V_i$ is {\it satisfied} if $d_{V_i}(v)\geq d_{V_{3-i}}(v) $, or 
equivalently if $d_{V_i}(v)\geq \lceil\frac{d(v)}{2} \rceil $. A graph admitting a non-trivial
partition where all vertices are satisfied is called {\it satisfactory partitionable}, and such a 
partition is called {\it satisfactory partition}. 
\par A problem with input size $n$ and parameter $k$ is said to be `fixed-parameter tractable (FPT)' if it has an algorithm that runs in time $\mathcal{O}(f(k)n^c)$, where $f$ is some (usually computable) function, and $c$ is a constant that does not depend on $k$ or $n$. What makes the theory more interesting is a hierarchy of intractable parameterized problem classes above FPT which helps in distinguishing those problems that are not fixed parameter tractable. Closely related to fixed-parameter tractability is the notion of preprocessing. 
%    The introduction of just one more dimension of analysis (the parameter) allows for the first time a useful mathematical theory of polynomial-time preprocessing to be developed: the field of  kernelization which is rapidly becoming a branch of theoretical computer science all in its own, with very strong and natural ties to practical computing and heuristics. 
    A reduction to a problem kernel, or equivalently, problem kernelization means to apply a data reduction process in polynomial time to an instance $(x, k)$ such that for the reduced instance $(x', k')$ it holds that $|x'| \leq g(k)$ and $k' \leq g(k)$ for some function $g$ only depending on $k$. Such a reduced instance is called a problem kernel. We refer to~\cite{marekcygan} for further details on parameterized complexity. 

\noindent {\it Our results:} Our main results are the following:
\begin{itemize}
    \item The {\sc  Satisfactory Partition} and {\sc Balanced  Satisfactory Partition} problems 
    are fixed parameter tractable (FPT) when parameterized by neighbourhood diversity. 
    \item The {\sc  Satisfactory Partition} and {\sc Balanced  Satisfactory Partition} problems  can be solved in polynomial time for graphs of bounded clique-width.
    \item A generalized version of the {\sc  Satisfactory Partition} problem is W[1]-hard when parameterized by treewidth. 
\end{itemize}

\noindent{\it Previous work:} In the first paper on this topic,  Gerber and Kobler  \cite{GERBER2000283} considered a 
generalized version of this problem by introducing weights for the vertices and edges and showed that
a general version of the problem is strongly NP-complete. For the unweighted version, they 
presented some sufficient conditions for the existence of a solution. This problem was further studied in 
\cite{isaac2003,TCScliquewidth,australia}. The {\sc Satisfactory Partition} problem is 
NP-complete and this implies that {\sc Balanced Satisfactory Partition} problem is also 
NP-complete via a simple reduction in which we add new dummy vertices and dummy edges to the 
graph \cite{cocoon05,BAZGAN20061236}.  Both problems are solvable in polynomial time 
for  graphs with maximum degree  at most 4 \cite{BAZGAN20061236}.  They also studied generalizations and variants of this problem when a partition into
$k \geq 3$ nonempty parts is required. Bazgan, Tuza, and Vanderpooten \cite{isaac2003,BAZGAN-TCS} studied 
an ``unweighted" generalization of {\sc Satisfactory Partition}, where each vertex $v$ 
is required to have at least $s(v)$ neighbours in its own part, for a given function 
$s$ representing the degree of satisfiability. Obviously, when 
$s=\lceil \frac{d}{2}\rceil$, where $d$ is the degree function, we obtain satisfactory
partition. They gave a polynomial-time algorithm for graphs of bounded treewidth which 
decides if a graph admits a satisfactory partition, and gives such a partition if it exists. \\

\section{FPT algorithm parameterized by neighbourhood diversity}\label{ndsection}

In this section, we present an FPT algorithm for the  {\sc Satisfactory Partition} and {\sc Balanced Satisfactory Partition} problems parameterized 
 by neighbourhood diversity. For a vertex $v\in V(G)$, we use $N_G(v)=\{u~:~(u,v)\in E(G)\}$ 
 to denote the (open) neighbourhood 
of vertex $v$ in $G$, and $N_G[v]=N_G(v)\cup \{v\}$ to denote the closed 
neighbourhood of $v$. The degree $d_G(v)$ of a vertex 
$v\in V(G)$ is $|N_G(v)|$. For a non-empty subset $S\subseteq V(G)$, we define its closed neighbourhood as $N_G[S]=\bigcup_{v\in S} N_G[v]$ and its 
open neighbourhood as $N_G(S)=N_G[S]\setminus S$. For a non-empty subset $A\subseteq V$ and 
a vertex $v\in V(G)$, $N_A(v)$ denotes the set of neighbours of $v$ in $A$, that is, 
$N_A(v)=\{ u\in A~:~ (u,v)\in E(G)\}$.  We use $d_A(v)=|N_A(v)|$ to denote the degree of vertex $v$ in $G[A]$,
the subgraph of $G$ induced by $A$. 
  We say two vertices $u$ and $v$ in $G$ have the same type if and only if 
 $N_G(u)\setminus \{v\}=N_G(v)\setminus \{u\}$. The relation of having the same type 
 is an equivalence  relation. The idea of neighbourhood diversity is based on this 
 type structure. 
  \begin{definition} \rm \cite{Lampis} 
        The neighbourhood diversity of a graph $G=(V,E)$, denoted by ${\tt nd}(G)$, is the least integer $k$ for which we can partition the set $V$ of vertices  into $k$ classes, such that all vertices in each class have the 
        same type.
    \end{definition}
     
    If neighbourhood diversity of a graph is bounded by an integer $k$, then there exists 
    a partition $\{ C_1, C_2,\ldots, C_k\}$ of $V(G)$ into $k$ type  classes. It is known that 
    such a minimum partition can be found in linear time using fast modular decomposition algorithms \cite{Tedder}. 
    %It is also known that ${\tt nd}(G) \leq 2^{{\tt vc}(G)}+{\tt vc}(G)$ for every graph $G$, where
    %{\tt vc}(G) denotes the size of a minimum vertex cover of $G$ \cite{Lampis}. 
    Notice
    that each type class  could either be a clique or an independent set by definition.  
     For algorithmic 
    purpose it is often useful to consider a {\it type graph} $H$ of  graph $G$, where
    each vertex of $H$ is a type class in $G$, and two vertices $C_i$ and $C_j$ are adjacent iff
    there is complete bipartite clique between these type classes in $G$. It is not difficult to see that
    there will be either a complete bipartite clique or no edges between any two type classes. 
     The key property of graphs of
  bounded neighbourhood diversity is that their type graphs have bounded size.  In this section, we prove the following theorem:
    \begin{theorem}\label{theoremnd1}
       The  {\sc Satisfactory Partition} problem  is fixed-parameter tractable when parameterized by  the neighbourhood diversity.
    \end{theorem}
    
 Let $G$ be a connected graph such that ${\tt nd}(G)=k$. Let $C_1,\ldots,C_k$ be the partition 
    of $V(G)$ into sets of type classes. We assume $k\geq 2$ since otherwise the problem 
     becomes trivial. 
    We define $I_1=\{ C_i~|~C_i\subseteq V_1\}$, $I_2=\{ C_i~|~C_i\subseteq V_2\}$ and 
    $I_3=\{ C_i~|~C_i\cap V_1\neq \emptyset, C_i\cap V_2, \neq \emptyset \}$ where $(V_1,V_2)$ is a 
    satisfactory partition. We next guess if $C_i$ belongs to $ I_1$, $I_2$, or $ I_3$.
    There are at most $3^k$ possibilities as each $C_i$ has three options: either in 
    $I_1$, $I_2$, or $I_3$. We reduce the problem of finding a satisfactory partition
    to an integer linear programming  optimization with $k$ variables. 
    Since integer linear programming is fixed parameter tractable when parameterized by 
    the number of variables \cite{lenstra}, we conclude that our problem is FPT when parameterized by 
    the neighbourhood diversity. \\

  \noindent {\bf  ILP Formulation:} Given $I_1, I_2$ and $I_3$, our goal here is to 
  answer if there exists a satisfactory partition $(V_1,V_2)$ of $G$ with 
 all vertices of  $C_i$  are in $V_1$ if $C_i\in I_1$, all vertices of $C_i$ are in $V_2$ if $C_i\in I_2$,
 and vertices of $C_i$ are distributed amongst  $V_1$ and  $V_2$ if $C_i\in I_3$. 
 For each $C_i$, we associate a variable: $x_i$ that indicates 
    $|V_1\cap C_i|=x_i$. Because the vertices in
    $C_i$ have the same neighbourhood, the variables $x_i$ determine $(V_1,V_2)$
    uniquely, up to isomorphism. 
    We now characterize a satisfactory partition in terms of $x_i$.
    Note that $x_i=n_i=|C_i|$  if $C_i\in I_1$; $x_i=0$  if $C_i\in I_2$. \\
    
\begin{lemma}
Let $C$ be a clique type class. Then $C$ is either in $I_1$ or $I_2$.
\end{lemma}  
\proof Let $C$ be a clique type class. Let $u,v\in C$ and  $N(u)\setminus \{v\}=N(v)\setminus \{u\}=\{w_1,\ldots, w_m\}$.
For the sake of contradiction, suppose $C$ is in $I_3$, that is, there exists a satisfactory
partition $(V_1,V_2)$ such 
that $u\in V_1$ and $v\in V_2$. Assume that $m$ is an even integer, that is,  $m=2\ell$ for
some integer $\ell$. As $u\in V_1$, at least 
$\ell+1$ vertices from the set $\{w_1,\ldots,w_{2\ell}\}$ must lie in $V_1$ in order 
to satisfy $u$. Similarly, as $v\in V_2$, at least 
$\ell+1$ vertices from the set $\{w_1,\ldots,w_{2\ell}\}$ must lie in $V_2$ in order 
to satisfy $v$. This is a contradiction, as there are only $2\ell$ vertices in the set.
A similar argument holds when $m$ is odd. This proves the lemma. \\

\noindent Now we consider the following four cases: \\
\noindent{\bf  Case 1:}  Suppose $v$ belongs to a clique type class $C_j$ in $I_1$. 
         Then the number of neighbours 
        of $v$ in $V_1$, that is, 
  $$d_{V_1}(v) =\sum\limits_{i:C_i\in N_H[C_j] \cap I_1}{n_i} +\Big(\sum\limits_{i:C_i\in N_H[C_j] \cap I_3}{x_i}\Big) -1.$$
 The number of neighbours of $v$ in $V_2$,  that is,  
   $$ d_{V_2}(v) =\sum\limits_{i: C_i\in N_H(C_j)\cap I_2}{n_i}+  \sum\limits_{i: C_i\in N_H[C_j] \cap I_3}{(n_i-x_i)}.$$
 Therefore, vertex  $v$  is satisfied if and only if  
 \begin{align}\label{eq1}
 \sum\limits_{i:C_i\in N_H[C_j] \cap I_1}{n_i} +\sum\limits_{i:C_i\in N_H[C_j] \cap I_3}{2x_i} & \geq 
1+ \sum\limits_{i:  C_i\in N_H(C_j)\cap I_2}{n_i}+  \sum\limits_{i:C_i\in N_H[C_j] \cap I_3}{n_i} 
 \end{align}
 
 \noindent{\bf  Case 2:}  Suppose $v$ belongs to a clique type  class $C_j$ in $I_2$. 
 Then similarly,  $v$  is satisfied  if and only if  
 \begin{align}\label{eq2}
  \sum\limits_{i:C_i\in N_H[C_j] \cap I_2}{n_i} +\sum\limits_{i:C_i\in N_H[C_j] \cap I_3}{n_i} & \geq 
 1+ \sum\limits_{i: C_i\in N_H(C_j)\cap I_1}{n_i}+  \sum\limits_{i:C_i\in N_H[C_j] \cap I_3}{2x_i}
 \end{align} 

\noindent{\bf  Case 3:}  Suppose $v$ belongs to an independent type class $C_j$ in $V_1$, 
that is, $C_j\in I_1\cup I_3$. 
         Then the number of neighbours 
        of $v$ in $V_1$, that is, 
 $$ d_{V_1}(v) =\sum\limits_{i:C_i\in N_H(C_j) \bigcap I_1}{n_i} +\sum\limits_{i:C_i\in N_H(C_j) \bigcap I_3}{x_i}. $$
          Note that if $C_j\in I_3$, then only $x_j$ 
        vertices of $C_j$ are in $V_1$ and the the remaining $y_j$ vertices of $C_j$ are in $V_2$. 
        The number of neighbours of $v$ in $V_2$,  that is,  
$$ d_{V_2}(v) =\sum\limits_{i: ~ C_i\in N_H(C_j)\bigcap I_2}{n_i}+  \sum\limits_{i:~C_i\in N_H(C_j) \bigcap I_3}{(n_i-x_i)}. $$
 Therefore,  $v$  is satisfied if and only if  
 \begin{align}\label{eq3}
  \sum\limits_{i:C_i\in N_H(C_j) \bigcap I_1}{n_i} +\sum\limits_{i:C_i\in N_H(C_j) \bigcap I_3}{2x_i}  & 
  \geq \sum\limits_{i: C_i\in N_H(C_j)\bigcap I_2}{n_i}+  \sum\limits_{i:C_i\in N_H(C_j) \bigcap I_3}{n_i}
 \end{align}
 
  \noindent{\bf  Case 4:}  Suppose $v$ belongs to an independent type class $C_j$ in $V_2$, that is, 
  $C_j\in I_2\cup I_3$. 
 Similarly,   vertex $v$  is satisfied if and only if  
 \begin{align}\label{eq4}
  \sum\limits_{i:C_i\in N_H(C_j) \bigcap I_2}{n_i} +\sum\limits_{i:C_i\in N_H(C_j) \bigcap I_3}{n_i} 
  \geq \sum\limits_{i:  C_i\in N_H(C_j)\bigcap I_1}{n_i}+  \sum\limits_{i:C_i\in N_H(C_j) \bigcap I_3}{2x_i}
 \end{align}

 We now formulate ILP formulation of satisfactory partition, for given $I_1, I_2$ and $I_3$. The question is whether 
 there exist $x_j$ under the conditions $x_j=n_j$ if $C_j\in I_1$, $x_j=0$ if $C_j\in I_2$,
 $x_j\in \{1,2,\ldots, n_j-1\}$ if 
 $ C_j\in I_3$ and the additional conditions 
 described below:
 \begin{itemize}
     \item Inequality \ref{eq1} for all clique type classes  $C_j\in I_1$
     \item Inequality \ref{eq2} for all clique type classes $C_j\in I_2$
     \item Inequality \ref{eq3} for all independent type classes $C_j\in I_1$
     \item Inequality \ref{eq4} for all independent type classes $C_j\in I_2$
     \item \begin{align*}
  \sum\limits_{C_i\in N_H(C_j) \bigcap I_2}{n_i} +\sum\limits_{C_i\in N_H(C_j) \bigcap I_3}{n_i} 
  = \sum\limits_{C_i\in N_H(C_j)\bigcap I_1}{n_i}+  \sum\limits_{C_i\in N_H(C_j) \bigcap I_3}{2x_i}
 \end{align*} for all independent type classes $C_j\in I_3$. 
 \end{itemize}
 For {\sc Balanced Satisfactory Partition} problem, we additionally ask that 
 $$\sum_{i:C_i\in I_1} {n_i} +\sum_{i:C_i\in I_3}{x_i}=\sum_{i:C_i\in I_3}{(n_i-x_i)}+\sum_{i:C_i\in I_2}{n_i}. $$

\noindent {\bf Solving the ILP:}
Lenstra \cite{lenstra} showed that the feasibility version of {\sc $p$-ILP} is FPT with 
running time doubly exponential in $p$, where $p$ is the number of variables. 
Later, Kannan \cite{kannan} designed an algorithm for {\sc $p$-ILP} running in time $p^{O(p)}$.\\

\noindent {\sc $p$-Variable Integer Linear Programming Feasibility ($p$-ILP)}: Let matrices $A\in \ Z^{m\times p}$ and $b\in \ Z^{p\times 1}$  be given. The question is whether there exists 
a vector $ x\in \ Z ^{p\times 1}$ satisfying  the $m$ 
inequalities, that is, $A\cdot x\leq b$. We use the following result:

\begin{lemma}\rm \label{ilp}\cite{lenstra,kannan,Frank10.1007/BF02579200}
{\sc $p$--ILP} can be solved using $O(p^{2.5p+o(p)}\cdot L)$ arithmetic operations and space polynomial in $L$. 
Here $L$ is the number of bits in the input. 
\end{lemma}

In the formulation for {\sc Satisfactory Partition} problem, we have at most $k$ variables. 
The value of any variable 
in the integer linear programming is  bounded by $n$, the number of vertices in the input graph. 
The constraints can be represented using 
$O(k^2 \log{n})$ bits. Lemma \ref{ilp} implies that we can solve the problem with the given
guess $I_1,I_2$ and $I_3$ in FPT time. 
There are at most $3^k$ choices for $(I_1,I_2,I_3$), and the ILP formula for a guess can be solved in FPT time. Thus 
Theorem \ref{theoremnd1} holds.

\section{Graphs of bounded  clique-width} 
This section presents a polynomial time 
algorithm for the {\sc Satisfactory Partition} and  {\sc Balanced Satisfactory Partition}
problems  for graphs of bounded clique-width. 
The clique-width of a graph 
$G$ is a parameter that describes the structural complexity of the graph; it is closely related to treewidth, 
but unlike treewidth it can be bounded even for dense graphs.   In a vertex-labeled graph, an $i$-vertex is a 
vertex of label $i$. \\
A $c$-expression is a rooted binary tree $T$ such that
\begin{itemize}
    \item each leaf has label $o_i$ for some $i\in \{1,\ldots,c\}$,
    \item each non-leaf node with two children has label $\cup $, and 
    \item each non-leaf node with only one child has label $\rho_{i,j}$ 
    or $\eta_{i,j}$ $(i,j\in \{1,\ldots,c\}, i\neq j)$.
\end{itemize}
Each node in a $c$-expression represents a vertex-labeled graph as follows: 
\begin{itemize}
    \item a $o_i$ node represents a graph with one $i$-vertex;
    \item a $\cup$-node represents the disjoint union of the labeled graphs represented 
    by its children;
    \item a $\rho_{ij}$-node represents the labeled graph obtained from the one
    represented by its child by replacing the labels of the
    $i$-vertices with $j$;
    \item a $\eta_{ij}$-node represents the labeled graph obtained from the one
    represented by its child by adding all possible edges between $i$-vertices
    and $j$-vertices.
\end{itemize}
A $c$-expression represents the graph represented by its root. A $c$-expression 
of a $n$-vertex graph $G$ has $O(n)$ vertices.  The {\it clique-width} of a 
graph $G$, denoted by ${\tt cw}(G)$, is the minimum $c$ for which there exists a $c$-expression $T$
representing a graph isomorphic to $G$. \\
A $c$-expression of a graph is {\it  irredundant}  if for each edge $\{u,v\}$, there is exactly 
one node $\eta_{i,j}$ that adds the edge between $u$ and $v$. It is known that a $c$-expression 
of a graph can be transformed into an irredundant one with $O(n)$ nodes in linear time \cite{COURCELLE200077}. 
Here we use irredundant $c$-expression only. 

Computing the clique-width and a corresponding $c$-expression of a graph is NP-hard \cite{10.1137/070687256}. 
For $c\leq 3$, we can compute a $c$-expression  of a graph of clique-width at most $c$
in $O(n^2m)$ time \cite{CORNEIL2012834}, where $n$ and $m$ are the number of vertices and edges, respectively. 
For fixed $c\geq 4$, it is not known whether one can compute the clique-width and a corresponding 
$c$-expression of a graph in polynomial time. On the 
other hand, it is known that for any fixed $c$, one can compute a 
$(2^{c+1}-1)$-expression of a graph of clique-width $c$
in $O(n^3)$ time \cite{oum}. For more details see \cite{KIYOMI201791}. Now we prove the following theorem. 

\begin{theorem}\label{cliqueth}
Given an $n$-vertex graph $G$ and an irredundant $c$-expression $T$ of $G$, the {\sc Satisfactory Partition} and {\sc Balanced Satisfactory
Partition} problems are solvable  in $O(n^{8c})$ time. 
\end{theorem}

For each node $t$ in a $c$-expression $T$, let $G_t$ be the vertex-labeled graph represented by $t$. 
We denote by $V_t$ the vertex set of $G_t$. For each $i$, we denote the set of 
$i$-vertices in $G_t$ by $V_t^i$. For each node $t$ in $T$, we construct a table $dp_t({\tt r,\bar{r}, s,\bar{s}}) \in \{\mbox{true, false}\} $
with indices ${\tt r}: \{1,\ldots,c\}\rightarrow \{0,\ldots,n\}$, 	${\tt \bar{r}}: \{1,\ldots,c\}\rightarrow \{0,\ldots,n\}$, 
 ${\tt s}: \{1,\ldots,c\}\rightarrow \{-n+1,\ldots,n-1\}\cup \{\infty\}$, and 
 ${\tt \bar{s}}: \{1,\ldots,c\}\rightarrow \{-n+1,\ldots,n-1\}\cup \{\infty\}$ 
as follows. We set $dp_t({\tt r,\bar{r}, s,\bar{s}})=\mbox{true} $ if and only if there exists a partition  $(S,\bar{S})$ of $ V_t$ such that 
for all $i\in\{1,2,\ldots,c\}$
\begin{itemize}
    \item ${\tt r}(i)=|S\cap V_t^i|$;
    \item ${\tt \bar{r}}(i)=|\bar{S}\cap V_t^i|$;
    \item if $S\cap V_t^i \neq \emptyset$, then 
    ${\tt {s}}(i)=\mbox{min}_{v\in S\cap V_t^i}\Big\{  |N_{G_t}(v)\cap S|-|N_{G_t}(v)\setminus S|\Big\}$, otherwise 
    ${\tt {s}}(i)=\infty$;
    \item if $\bar{S}\cap V_t^i \neq \emptyset$, then 
    ${\tt \bar{s}}(i)=\mbox{min}_{v\in \bar{S}\cap V_t^i}\Big\{  |N_{G_t}(v)\cap \bar{S}|-|N_{G_t}(v)\setminus \bar{S}|\Big\}$, otherwise 
    ${\tt \bar{s}}(i)=\infty$.    
\end{itemize}
That is, ${\tt r}(i)$ denotes the number of the $i$-vertices in $S$;  ${\tt \bar{r}}(i)$ denotes the number of the $i$-vertices
 in $\bar{S}$;  ${\tt {s}}(i)$ is the ``surplus" at the weakest $i$-vertex in $S$ and  ${\tt \bar{s}}(i)$ is the ``surplus" at the weakest
  $i$-vertex in $\bar{S}$. 
\par Let $\tau$ be the root of the $c$-expression $T$ of $G$. Then  $G$ has a  satisfactory partition 
if there exist ${\tt r,\bar{r}, s,\bar{s}} $ satisfying 
\begin{enumerate}
    \item $dp_{\tau}({\tt r,\bar{r}, s,\bar{s}})=\mbox{true} $;
    \item $\mbox{min}\Big\{ {\tt s}(i), {\tt\bar{s}}(i)\Big\}\geq 0$. 
\end{enumerate} 
For the {\sc Balanced Satisfactory Partition} problem, we additionally ask that $\sum_{i=1}^c {{\tt r}(i)} =\sum_{i=1}^c{{\tt \bar{r}}(i)}$.
If all entries $dp_{\tau}({\tt r,\bar{r}, s,\bar{s}})$ are computed 
in advance, then we can  verify above conditions by spending $O(1)$ time 
for each tuple $({\tt r,\bar{r}, s,\bar{s}})$. 
 \par In the following, we compute all entries $dp_{t}({\tt r,\bar{r}, s,\bar{s}})$
in a bottom-up manner. There are $(n+1)^c \cdot (n+1)^c\cdot (2n)^c \cdot (2n)^c=$ $O(n^{4c})$ 
possible tuples $({\tt r,\bar{r}, s,\bar{s}})$. Thus, to prove Theorem \ref{cliqueth}, it is enough to prove that 
each entry $dp_{t}({\tt r,\bar{r}, s,\bar{s}})$ can be computed in time $O(n^{4c})$ assuming that 
the entries for the children of $t$ are already computed.

\begin{lemma}\rm 
For a leaf node $t$ with label $o_i$, $dp_t({\tt r,\bar{r}, s,\bar{s}})$ can be computed in $O(1)$ time. 
\end{lemma}
\proof Observe that $dp_t({\tt r,\bar{r}, s,\bar{s}})={\tt true}$ if and only if 
${\tt r}(j)=0$, ${\tt \bar{r}}(j)=0$, ${\tt s}(j)=0$, and ${\tt \bar{s}}(j)=0$ for all $j\neq i$ and either 
  \begin{itemize}  
  \item   ${\tt r}(i)=0$, ${\tt \bar{r}}(i)=1$, ${\tt s}(i)=\infty$, ${\tt \bar{s}}(i)=0$, or 
 \item  ${\tt r}(i)=1$, ${\tt \bar{r}}(i)=0$, ${\tt s}(i)=0$, ${\tt \bar{s}}(i)=\infty$.  
 \end{itemize}
The first case corresponds to $S=\emptyset, \bar{S}=V_t^i$, and the second case 
corresponds to $S=V_t^i, \bar{S}=\emptyset$. These conditions can be checked in $O(1)$ time. 

\begin{lemma}\rm 
For a $\cup$-node $t$, $dp_t({\tt r,\bar{r}, s,\bar{s}})$ can be computed in $O(n^{4c})$ time.
\end{lemma}

\proof Let $t_1$ and $t_2$ be the children of $t$ in $T$. Then $dp_t({\tt r,\bar{r}, s,\bar{s}})={\tt true}$ 
if and only if there exist ${\tt r_1,\bar{r}_1, s_1,\bar{s}_1}$ and ${\tt r_2,\bar{r}_2, s_2,\bar{s}_2}$ such that 
$dp_t({\tt r_1,\bar{r}_1, s_1,\bar{s}_1})={\tt true}$,  $dp_t({\tt r_2,\bar{r}_2, s_2,\bar{s}_2})={\tt true}$,
${\tt r}(i)={\tt r_1}(i)+{\tt r_2}(i)$,  ${\tt \bar{r}}(i)={\tt \bar{r}}_1(i)+{\tt \bar{r}}_2(i)$, 
${\tt s}(i)=\mbox{min}\Big\{ {\tt s}_1(i), {\tt s}_2(i)\Big\}$ and 
${\tt \bar{s}}(i)=\mbox{min}\Big\{ {\tt \bar{s}}_1(i), {\tt \bar{s}}_2(i)\Big\}$ for all $i$. The number of possible pairs for 
${\tt (r_1,r_2)}$ is  at most $(n+1)^c$ as ${\tt r_2}$ is uniquely determined by
${\tt r_1}$; the number of possible pairs for 
${\tt (\bar{r}_1,\bar{r}_2)}$ is  at most $(n+1)^c$ as ${\tt \bar{r}_2}$ is uniquely determined by
${\tt \bar{r}_1}$.  There are at most $2^c(2n)^c$ possible pairs for $({\tt s_1,s_2})$ and for 
$({\tt \bar{s}_1,\bar{s}_2})$ each.  In total, there are $O(n^{4c})$ candidates. Each candidate can be checked in 
$O(1)$ time, thus the lemma holds. 

\begin{lemma}\rm 
For a $\eta_{ij}$-node $t$, $dp_t({\tt r,\bar{r}, s,\bar{s}})$ can be computed in $O(1)$ time.
\end{lemma}

\proof Let $t^{\prime}$ be the child of $t$ in $T$. Then,  
$dp_t({\tt r,\bar{r}, s,\bar{s}})={\tt true}$ if and only if \\
$dp_t({\tt r,\bar{r}, s^{\prime},\bar{s}^{\prime}})={\tt true}$ 
for some ${\tt s^{\prime},\bar{s}^{\prime}}$ with the following conditions:
\begin{itemize}
    \item ${\tt s}(h)={\tt s}^{\prime}(h)$ and ${\tt \bar{s}}(h)={\tt \bar{ s}}^{\prime}(h)$ hold for all $h\notin \{i,j\}$;
    \item ${\tt s}(i)={\tt s^{\prime}}(i)+2{\tt r}(j) -|V_t^j|$ and  ${\tt s}(j)={\tt s^{\prime}}(j)+2{\tt r}(i) -|V_t^i|$;
    \item  ${\tt \bar{s}}(i)={\tt \bar{s}^{\prime}}(i)+2{\tt \bar{r}}(j) -|V_t^j|$ and  ${\tt \bar{s}}(j)={\tt \bar{s}^{\prime}}(j)+2{\tt \bar{r}}(i) -|V_t^i|$. 
    \end{itemize}
 We now explain the condition for $s(i)$. Recall that $T$ is irredundant. That is, the graph $G_{t^{\prime}}$ does not have 
any edge between the $i$-vertices and the $j$-vertices. In $G_t$, an $i$-vertex has exactly
${\tt r}(j)$ more neighbours in $S$ and exactly $|V_t^j|-{\tt r}(j)$ more neighbours in $\bar{S}$.
Thus we have ${\tt s}(i)={\tt s^{\prime}}(i)+2{\tt r}(j) -|V_t^j|$. The  lemma holds as there is only one candidate for each 
${\tt s}^{\prime}(i)$, ${\tt s}^{\prime}(j)$, ${\tt \bar{s}}^{\prime}(i)$ and 
${\tt \bar{s}}^{\prime}(j)$.

\begin{lemma}\rm 
For a $\rho_{ij}$-node $t$, $dp_t({\tt r,\bar{r}, s,\bar{s}})$ can be computed in $O(n^4)$ time.
\end{lemma}

\proof Let $t^{\prime}$ be the child of $t$ in $T$. Then,  $dp_t({\tt r,\bar{r}, s,\bar{s}})={\tt true}$ 
if and only if there exist  ${\tt r}^{\prime}, {\tt \bar{r}}^{\prime}, {\tt s}^{\prime}, {\tt \bar{s}}^{\prime}$ such that 
$dp_{t^{\prime}}({\tt r^{\prime},\bar{r}^{\prime}}, {\tt s}^{\prime}, {\tt \bar{s}}^{\prime})
={\tt true}$, where :
\begin{itemize}
\item ${\tt r}(i)=0 $, ${\tt r}(j)={\tt r}^{\prime}(i)+{\tt r}^{\prime}(j)$, and 
${\tt r}(h)={\tt r}^{\prime}(h)$ if $h\notin \{i,j\}$;
\item ${\tt \bar{r}}(i)=0 $, ${\tt \bar{r}}(j)={\tt \bar{r}}^{\prime}(i)+{\tt \bar{r}}^{\prime}(j)$, and 
${\tt\bar{ r}}(h)={\tt \bar{r}}^{\prime}(h)$ if $h\notin \{i,j\}$;
\item ${\tt s}(i)=\infty $, ${\tt s}(j)=\mbox{min}\big\{{\tt s}^{\prime}(i), {\tt s}^{\prime}(j)\big\}$, and 
${\tt s}(h)={\tt s}^{\prime}(h)$ if $h\notin \{i,j\}$;
\item ${\tt \bar{s}}(i)=\infty $, ${\tt \bar{s}}(j)=\mbox{min}\big\{{\tt \bar{s}}^{\prime}(i), {\tt \bar{s}}^{\prime}(j)\big\}$, and 
${\tt \bar{s}}(h)={\tt \bar{s}}^{\prime}(h)$ if $h\notin \{i,j\}$.
\end{itemize}
The number of possible pairs for $({\tt r}^{\prime}(i), {\tt r}^{\prime}(j))$ is $O(n)$ as 
${\tt r}^{\prime}(j)$ is uniquely determined by ${\tt r}^{\prime}(i)$; similarly the number of possible pairs for 
$({\tt \bar{r}}^{\prime}(i), {\tt \bar{r}}^{\prime}(j))$ is $O(n)$ as 
${\tt \bar{r}}^{\prime}(j)$ is uniquely determined by ${\tt \bar{r}}^{\prime}(i)$. 
There are at most $O(n)$ possible pairs for $({\tt s^{\prime}}(i),{\tt s^{\prime}}(j))$ and for 
$({\tt \bar{s}^{\prime}}(i),{\tt \bar{s}^{\prime}}(j))$.  In total, there are $O(n^{4})$ candidates. Each candidate can be checked in 
$O(1)$ time, thus the lemma holds. \\

 \section{W[1]-hardness parameterized by treewidth} 
  In this section we show that a generalization of {\sc Satisfactory Partition} is W[1]-hard when parameterized by treewidth.
 We consider the following generalization of {\sc Satisfactory partition}, where
 some vertices are forced to be in the first part $V_1$
 and some other vertices are
 forced to be in the second part $V_2$. 
 \vspace{3mm}
    \\
    \fbox
    {\begin{minipage}{33.7em}\label{SP2}
       {\sc  Satisfactory Partition$^{\mbox{FS}}$}\\
        \noindent{\bf Input:} A graph $G=(V,E)$, a set $V_{\triangle}\subseteq V(G)$, and
        a set $V_{\square}\subseteq V(G)$. \\
        %, and 
        %a set $C\subseteq V(G)\times V(G)$.\\
    \noindent{\bf Question:} Is there a satisfactory partition $(V_1,V_2)$ of $V$
    such that (i) $V_{\triangle}\subseteq V_1$ (ii) $V_{\square}\subseteq V_2$.
    %, and (iii) for all $(a,b)\in C$, $V_1$ 
    %contains either $a$ or $b$ but not both?
    \end{minipage} }\\
    
    \noindent In this section, we prove the following theorem:
    \begin{theorem}\label{twtheorem}
    The {\sc  Satisfactory Partition$^{\mbox{FS}}$} is W[1]-hard when parameterized by the treewidth of the 
    graph.
     \end{theorem}
     
     We prove W[1]-hardness by using techniques from a hardness result for the problem of finding minimum 
     defensive alliance in a graph \cite{BLIEM2018334}. 
 Let $G=(V,E)$ be an undirected and edge weighted graph, where $V$, $E$, and $w$ denote
 the set of nodes, the set of edges and a positive integral weight 
 function $w:~E\rightarrow Z^{+}$, respectively. An orientation  $\Lambda$ of $G$ is an
 assignment of a direction to each edge $\{u,v\}\in E(G)$, that is, either $(u,v)$ or $(v,u)$
 is contained in $\Lambda$. The weighted outdegree of $u$ on $\Lambda$ is $w_{\mbox{out}}^u=\sum_{(u,v)\in \Lambda}w(\{u,v\})$.
 We define {\sc  Minimum Maximum Outdegree} problem as follows: 
 \vspace{3mm}
    \\
    \fbox
    {\begin{minipage}{33.7em}\label{SP3}
       {\sc  Minimum Maximum Outdegree}\\
        \noindent{\bf Input:} A graph $G$, an edge weighting $w$ of $G$ given in unary, and a positive integer $r$. \\
    \noindent{\bf Question:} Is there an orientation $\Lambda$  of $G$ such that $w_{\mbox{out}}^u\leq r$ for 
    each $u\in V(G)$?
    \end{minipage} }\\
    
\noindent It is known that {\sc  Minimum Maximum Outdegree} is W[1]-hard when parameterized by the treewidth of the input graph \cite{DBLP:journals/corr/abs-1107-1177}. We reduce this problem to the following generalization of {\sc Satisfactory Partition } problem:
\vspace{3mm}
    \\
    \fbox
    {\begin{minipage}{33.7em}\label{SP4}
       {\sc  Satisfactory Partition$^{\mbox{FSC}}$}\\
        \noindent{\bf Input:} A graph $G=(V,E)$, a set $V_{\triangle}\subseteq V(G)$, 
        a set $V_{\square}\subseteq V(G)$, and a set $C\subseteq V(G)\times V(G)$.\\
    \noindent{\bf Question:} Is there a satisfactory partition $(V_1,V_2)$ of $V$
    such that (i) $V_{\triangle}\subseteq V_1$ (ii) $V_{\square}\subseteq V_2$, and (iii) for all $(a,b)\in C$, $V_1$ 
    contains either $a$ or $b$ but not both?
    \end{minipage} }\\
    
   To prove Theorem \ref{twtheorem}, we give a 2-step reduction. In the first step 
   of the reduction, we  reduce {\sc Minimum Maximum Outdegree} to  {\sc  Satisfactory Partition$^{\mbox{FSC}}$}.
   In the second step of the reduction we reduce the {\sc  Satisfactory Partition$^{\mbox{FSC}}$}  to {\sc  Satisfactory Partition$^{\mbox{FS}}$}.
   To measure the treewidth  of a {\sc  Satisfactory Partition$^{\mbox{FSC}}$} instance, we use the following definition.
   Let $I=(G, V_{\triangle}, V_{\square}, C)$ be a {\sc  Satisfactory Partition$^{\mbox{FSC}}$} instance. The 
   {\it  primal graph} $G^{\prime}$ of $I$ is defined as follows: $V(G^{\prime})=V(G)$ and $E(G^{\prime})=E(G)\cup C$.
 
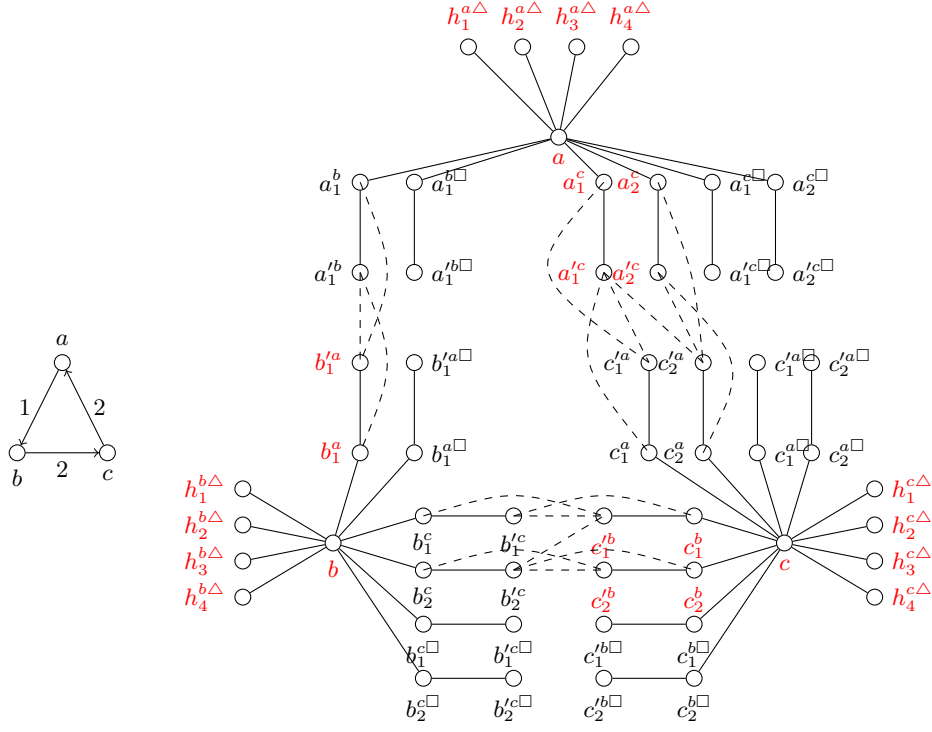
\begin{figure}[ht]
     \centering
    \[\begin{tikzpicture}[scale=1.2]
	%% Notice in the first vertex is named (v) for the sake of a later edge,
	%% and it also has a label to its left that is the math-mode $v$. 
	\vertex (b) at (.5,0) [label=below:${\color{red} b}$] {};
	\vertex (hb1) at (-0.5,0.6) [label=left:${\color{red} h^{b\triangle}_1}$] {}; 
	\vertex (hb2) at (-0.5, 0.2) [label=left:${\color{red} h^{b\triangle}_2}$] {}; 
	\vertex (hb3) at (-0.5,-0.2) [label=left:${\color{red} h^{b\triangle}_3}$] {}; 
	\vertex (hb4) at (-0.5,-0.6) [label=left:${\color{red} h^{b\triangle}_4}$] {};
	\vertex (bc11) at (1.5,0.3) [label=below:${ b^c_1}$] {}; 
	\vertex (bc12) at (1.5, -0.3) [label=below:${ b^c_2}$] {}; 
	\vertex (bc13) at (1.5,-0.9) [label=below:$b^{c \square}_1$] {}; 
	\vertex (bc14) at (1.5,-1.5) [label=below:$b^{c \square}_2$] {}; 
	\vertex (bc21) at (2.5,0.3) [label=below:${ b^{\prime c}_1}$] {}; 
	\vertex (bc22) at (2.5, -0.3) [label=below:${ b^{\prime c}_2}$] {}; 
	\vertex (bc23) at (2.5,-0.9) [label=below:${ b^{\prime c \square}_1}$] {}; 
	\vertex (bc24) at (2.5,-1.5) [label=below:${ b^{\prime c \square}_2}$] {}; 
	\vertex (ba11) at (0.8,1) [label=left:${\color{red} b^a_1}$] {}; 
	\vertex (ba13) at (1.4,1) [label=right:$b^{a \square}_1$] {}; 
	\vertex (ba21) at (0.8,2) [label=left:${\color{red} b^{\prime a}_1}$] {}; 
	\vertex (ba23) at (1.4,2) [label=right:${ b^{\prime a \square}_1}$] {}; 
	\vertex (c) at (5.5,0) [label=below:$\color{red} c$] {};
	\vertex (hc1) at (6.5,0.6) [label=right:${\color{red} h^{c\triangle}_1}$] {}; 
	\vertex (hc2) at (6.5, 0.2) [label=right:${\color{red} h^{c\triangle}_2}$] {}; 
	\vertex (hc3) at (6.5,-0.2) [label=right:${\color{red} h^{c\triangle}_3}$] {}; 
	\vertex (hc4) at (6.5,-0.6) [label=right:${\color{red} h^{c\triangle}_4}$] {}; 
	\vertex (cb11) at (4.5,0.3) [label=below:${\color{red} c^b_1}$] {}; 
	\vertex (cb12) at (4.5, -0.3) [label=below:${\color{red} c^b_2}$] {}; 
	\vertex (cb13) at (4.5,-0.9) [label=below:$c^{b \square}_1$] {}; 
	\vertex (cb14) at (4.5,-1.5) [label=below:$c^{b \square}_2$] {}; 
	\vertex (cb21) at (3.5,0.3) [label=below:${\color{red} c^{\prime b}_1}$] {}; 
	\vertex (cb22) at (3.5, -0.3) [label=below:${\color{red} c^{\prime b}_2}$] {}; 
	\vertex (cb23) at (3.5,-0.9) [label=below:${ c^{\prime b \square}_1}$] {}; 
	\vertex (cb24) at (3.5,-1.5) [label=below:${ c^{\prime b \square}_2}$] {}; 
	\vertex (ca11) at (4,1) [label=left:${ c^a_1}$] {}; 
	\vertex (ca12) at (4.6,1) [label=left:${ c^a_2}$] {}; 
	\vertex (ca13) at (5.2,1) [label=right:$c^{a \square}_1$] {}; 
	\vertex (ca14) at (5.8,1) [label=right:$c^{a \square}_2$] {}; 
	\vertex (ca21) at (4,2) [label=left:${ c^{\prime a}_1}$] {}; 
	\vertex (ca22) at (4.6, 2) [label=left:${ c^{\prime a}_2}$] {}; 
	\vertex (ca23) at (5.2,2) [label=right:${ c^{\prime a \square}_1}$] {}; 
	\vertex (ca24) at (5.8,2) [label=right:${ c^{\prime a \square}_2}$] {}; 
	\vertex (a) at (3,4.5) [label=below:$\color{red} a$] {};
	\vertex (ac11) at (3.5,4) [label=left:${\color{red} a^c_1}$] {}; 
	\vertex (ac12) at (4.1,4) [label=left:${\color{red} a^c_2}$] {}; 
	\vertex (ac13) at (4.7,4) [label=right:$a^{c \square}_1$] {}; 
	\vertex (ac14) at (5.4,4) [label=right:$a^{c \square}_2$] {}; 
	\vertex (ac21) at (3.5,3) [label=left:${\color{red} a^{\prime c}_1}$] {}; 
	\vertex (ac22) at (4.1,3) [label=left:${\color{red} a^{\prime c}_2}$] {}; 
	\vertex (ac23) at (4.7,3) [label=right:${ a^{\prime c \square}_1}$] {}; 
	\vertex (ac24) at (5.4,3) [label=right:${ a^{\prime c \square}_2}$] {}; 
	\vertex (ab11) at (0.8,4) [label=left:${ a^b_1}$] {}; 
	\vertex (ab13) at (1.4,4) [label=right:$a^{b \square}_1$] {}; 
	\vertex (ab21) at (0.8,3) [label=left:${ a^{\prime b}_1}$] {}; 
	\vertex (ab23) at (1.4,3) [label=right:${ a^{\prime b \square}_1}$] {}; 
	\vertex (ha1) at (2,5.5) [label=above:$\color{red} h^{a\triangle}_1$] {};
	\vertex (ha2) at (2.6,5.5) [label=above:$\color{red} h^{a\triangle}_2$] {};
	\vertex (ha3) at (3.2,5.5) [label=above:$\color{red} h^{a\triangle}_3$] {};
	\vertex (ha4) at (3.8,5.5) [label=above:$\color{red} h^{a\triangle}_4$] {};
%	\vertex (y) at (1,0.5) [label=below:$\square^{ab}$] {};
	\path
	   % Note that the word "path" here isn't used in the graph-theory sense; the \path command
	   % is always used prior to the list of edges; here, coincidentally, they do form an actual path.
	   (a) edge (ha1)
	   (a) edge (ha2)
	   (a) edge (ha3)
	   (a) edge (ha4)
	   (a) edge (ac11)
	   (a) edge (ac12)
	   (a) edge (ac13)
	   (a) edge (ac14)
	   (ac11) edge (ac21)
       (ac12) edge (ac22)
       (ac13) edge (ac23)
       (ac14) edge (ac24)
       (a) edge (ab11)
	   (a) edge (ab13)
	   (ab11) edge (ab21)
       (ab13) edge (ab23)
	   (b) edge (hb1)
	   (b) edge (hb2)
	   (b) edge (hb3)
	   (b) edge (hb4)
	   (b) edge (bc11)
	   (b) edge (bc12)
	   (b) edge (bc13)
	   (b) edge (bc14)
	   (bc11) edge (bc21)
       (bc12) edge (bc22)
       (bc13) edge (bc23)
       (bc14) edge (bc24)
       (b) edge (ba11)
	   (b) edge (ba13)
	   (ba11) edge (ba21)
       (ba13) edge (ba23)
	   (c) edge (hc1)
	   (c) edge (hc2)
	   (c) edge (hc3)
	   (c) edge (hc4)
	   (c) edge (cb11)
	   (c) edge (cb12)
	   (c) edge (cb13)
	   (c) edge (cb14)
	   (cb11) edge (cb21)
       (cb12) edge (cb22)
       (cb13) edge (cb23)
       (cb14) edge (cb24)
       (c) edge (ca11)
	   (c) edge (ca12)
	   (c) edge (ca13)
	   (c) edge (ca14)
	   (ca11) edge (ca21)
       (ca12) edge (ca22)
       (ca13) edge (ca23)
       (ca14) edge (ca24)
	   ;
       
       %\draw (-2,2) .. controls (-1,0) and (1,0) .. (2,2);
       %\path[draw, dashed] (cb21) -- controls (3,0.5) -- (bc21);
       \draw[dashed] (2.5,0.3) -- (3.5,0.3);
       \draw[dashed] (2.5,-0.3) -- (3.5,-0.3);
       \draw[dashed] (2.5,-0.3) -- (3.5,0.3);
       \draw[dashed] (2.5,0.3) .. controls (3.5,0.6) .. (4.5,0.3);
       \draw[dashed] (2.5,-0.3) .. controls (3.5,0) .. (4.5,-0.3);
       \draw[dashed] (1.5,0.3) .. controls (2.5,0.6) .. (3.5,0.3);
       \draw[dashed] (1.5,-0.3) .. controls (2.5,0) .. (3.5,-0.3);
       \draw[dashed] (3.5,3) -- (4,2);
       \draw[dashed] (4.1,3) -- (4.6,2);
       \draw[dashed] (3.5,3) -- (4.6,2);
       \draw[dashed] (3.5,4) .. controls (2.6,3) .. (4,2);
       \draw[dashed] (3.5,3) .. controls (3.2,2) .. (4,1);
       \draw[dashed] (4.1,4) .. controls (4.5,3) .. (4.6,2);
       \draw[dashed] (4.1,3) .. controls (5.1,2) .. (4.6,1);
       \draw[dashed] (0.8,3) -- (0.8,2);
       \draw[dashed] (0.8,4) .. controls (1.2,3) .. (0.8,2);
       \draw[dashed] (0.8,3) .. controls (1.2,2) .. (0.8,1);

	\vertex (a) at (-2.5, 2) [label=above:$a$]{};
	\vertex (b) at (-3, 1) [label=below:$b$]{};
	\vertex (c) at (-2, 1) [label=below:$c$]{};
	\path[->]
		(c) edge node[right]{$2$}  (a)
		(b) edge node[below]{$2$} (c)
		(a) edge node[left]{$1$} (b)
		;
 \end{tikzpicture}\]
\caption{Result of our reduction on a {\sc Minimum Maximum Outdegree} instance $G$ with $r=2$. The graph $G$ is shown at the left; and $G^{\prime}$ is shown at the right. Complementary vertex pairs are shown using dashed lines. The vertices in the first part
 of satisfactory partition $(V_1,V_2)$ of $G^{\prime}$
are shown in red for the given orientation of $G$.}
     \label{satfig1}
 \end{figure}

    \begin{lemma}
    The {\sc  Satisfactory Partition$^{\mbox{FSC}}$} is W[1]-hard when parameterized by the treewidth of the 
    primal graph.
    \end{lemma}
    \proof Let  $G=(V,E,w)$ and a positive integer $r$ be an  instance of {\sc Minimum Maximum Outdegree}.
    We construct an instance of {\sc  Satisfactory Partition$^{\mbox{FSC}}$} as follows. An example is given in Figure 
    \ref{satfig1}.
    For each vertex $v\in V(G)$, we introduce a set of new vertices $H_v=\{h_1^{v\triangle}, \dots, h_{2r}^{v\triangle}\}$. For each edge 
    $(u,v)\in E(G)$, we introduce the set of new vertices 
    $V_{uv}=\{u^v_1, \ldots, u^v_{w(u,v)}\}$, $V^{\prime}_{uv}=\{{u^{\prime v}_1}, \ldots, u^{\prime v}_{w(u,v)}\}$,
    $V_{vu}=\{v^u_1, \ldots, v^u_{w(u,v)}\}$, $V^{\prime}_{vu}=\{{v^{\prime u}_1}, \ldots, v^{\prime u}_{w(u,v)}\}$,
     $V^{\square}_{uv}=\{u^{v\square}_1, \ldots, u^{v\square}_{w(u,v)}\}$, $V^{\prime\square}_{uv}=\{{u^{\prime v\square}_1}, \ldots, u^{\prime v\square}_{w(u,v)}\}$,
    $V^{\square}_{vu}=\{v^{ u\square}_1, \ldots, v^{u\square}_{w(u,v)}\}$, 
    $V^{\prime\square}_{vu}=\{{v^{\prime u\square}_1}, \ldots, v^{\prime u\square}_{w(u,v)}\}$. We now define the graph  
    $G^{\prime}$ with  
    \begin{align*}
    V(G^{\prime})&=V(G) \bigcup\limits_{v\in V(G)}H_v \bigcup\limits_{(u,v)\in E(G)}(V_{uv}\cup 
    V_{uv}^{\square} \cup V_{vu}\cup V_{vu}^{\square})\\
    & \bigcup\limits_{(u,v)\in E(G)}(V^{\prime}_{uv}\cup 
    V_{uv}^{\prime\square} \cup V^{\prime}_{vu}\cup V_{vu}^{\prime\square})
    \end{align*}
    and 
    \begin{align*}
 E(G^{\prime})&= \big\{(v,h)~|~v\in V(G), h\in H_v\big\} \\
      & \bigcup \Big\{ (u,x) ~|~ (u,v)\in E(G), x\in V_{uv}\cup V_{uv}^{\square}\Big\}\\
      & \bigcup \Big\{(x,v)~|~ (u,v)\in E(G), x \in V_{vu}\cup V_{vu}^{\square}\Big\}\\
      & \bigcup \Big\{ (u_i^v,u_i^{\prime v}), (u_i^{v\square},u_i^{\prime v\square}), (v_i^u,v_i^{\prime u}),(v_i^{u\square},v_i^{\prime u\square})~|~(u,v)\in E(G), 1\leq i \leq w(u,v)\Big\}.
\end{align*}
We define the complementary vertex pairs 
\[C=\Big\{(u_i^{\prime v}, v_i^{\prime u}),(u_{i+1}^{\prime v}, v_i^{\prime u}),
(u_i^{v}, v_i^{\prime u}), (u_i^{\prime v}, v_i^{u})~|~ (u,v)\in E(G), 1\leq i\leq w(u,v)\Big\}\]
Complementary vertex pairs are shown in dashed lines in Figure \ref{satfig1}. 
 Finally we define $V_{\triangle}=V(G)\bigcup_{v\in V(G)}H_v$  and 
 $V_{\square}=\bigcup_{(u,v)\in E(G)}(V_{uv}^{\square} \cup V_{uv}^{\prime\square}\cup V_{vu}^{\square}\cup V_{vu}^{\prime\square})$. We use $I$ to denote $(G^{\prime}, V_{\triangle}, V_{\square},C)$ which is an instance 
 of {\sc  Satisfactory Partition$^{\mbox{FSC}}$}.\\
 \par Clearly, it takes polynomial time to compute $I$. We now prove that  the treewidth of the primal 
 graph $G^{\prime}$ of $I$ is bounded by a function of  the treewidth of $G$. We do so by modifying an optimal tree decomposition $\tau$ of $G$ as follows:
 \begin{itemize}
     \item For each $(u,v)\in E(G)$, we take an arbitrary node whose bag $B$ contains 
     both $u$ and $v$ and add to it a chain of nodes $1,2,\ldots,w(u,v)-1$ such that the bag of node $i$
     is $B\cup \{u_i^v, u_i^{\prime v},  v_{i}^{\prime u}, v_{i}^{u}, u_{i+1}^{ v}, u_{i+1}^{\prime v}, v_{i+1}^{\prime u},
     v_{i+1}^{ u} \}$. 
     \item For each $(u,v)\in E(G)$, we take an arbitrary node whose bag $B$ contains 
      $u$ and add to it a chain of nodes $1,2,\ldots,w(u,v)$ such that the bag of node $i$
     is $B\cup \{u_i^{v \square}, u_i^{\prime v \square}\}$. 
     \item For each $(u,v)\in E(G)$, we take an arbitrary node whose bag $B$ contains 
      $v$ and add to it a chain of nodes $1,2,\ldots,w(u,v)$ such that the bag of node $i$
     is $B\cup \{v_i^{u \square}, v_i^{\prime u \square}\}$. 
     \item For each $v\in V(G)$, we take an arbitrary node whose bag $B$ contains 
      $v$ and add to it a chain of nodes $1,2,\ldots,2r$ such that the bag of node $i$
     is $B\cup \{h_i^{v\triangle}\}$. 
     
 \end{itemize}
 Clearly, the modified tree decomposition is a valid tree decomposition of the primal graph of $I$ and 
 its width is at most the treewidth of $G$ plus eight. 

Let $D$ be the directed graph obtained by  an orientation of the edges of $G$ such that for 
each vertex the sum of the weights of outgoing edges is at most $r$. Consider the partition 
$$V_1=V_{\triangle} \bigcup_{(u,v)\in E(D)}(V_{vu}\cup V_{vu}^{\prime})=V(G)\bigcup_{v\in V(G)}H_v \bigcup_{(u,v)\in E(D)}(V_{vu}\cup V_{vu}^{\prime})$$ and 
$$V_2=\bigcup_{(u,v)\in E(D)}(V_{uv}\cup V^{\prime}_{uv}\cup V_{uv}^{\square}\cup V^{\prime\square}_{uv} )
\bigcup_{(u,v)\in E(D)}(V_{vu}^{\square}\cup V^{\prime\square}_{vu}).$$ 
To prove that $(V_1,V_2)$ is a satisfactory partition, first we prove that 
$d_{V_1}(x)\geq d_{V_2}(x)$ for all $x\in V_1$. If $x$ is a vertex in $H_v$ or 
$V_{vu}\cup V_{vu}^{\prime}$, then clearly all neighbours of $x$ are in $V_1$, hence $x$ is
satisfied. Suppose $x\in V(G)$. Let $w_{\mbox{out}}^x$ and $w_{\mbox{in}}^x$ denote the sum of the weights 
of outgoing and incoming edges of vertex $x$, respectively. Hence $d_{V_1}(x)=2r+w_{\mbox{in}}^x$ and 
$d_{V_2}(x)=2w_{\mbox{out}}^x+w_{\mbox{in}}^x$ in $G^{\prime}$. This shows that $x$ is satisfied as 
$w_{\mbox{out}}^x\leq r$. Now we prove that $d_{V_2}(x)\geq d_{V_1}(x)$ for all $x\in V_2$. 
If $x$ is a vertex in $V_{uv}\cup V^{\square}_{uv}\cup V^{\square}_{vu}$ then $x$ has one neighbour in 
$V_1$ and one neighbour in $V_2$. If $x\in V^{\prime}_{uv}\cup V^{\prime\square}_{uv}\cup V^{\prime\square}_{vu}$ then $x$ has one neighbour in 
$V_2$ and no neighbours in $V_1$. Thus the vertices in $V_2$ are satisfied. 

\par Conversely, suppose $(V_1,V_2)$ is a satisfactory partition of $I$. 
For every $(u,v)\in E(G)$, either
$V_{uv}\cup V^{\prime}_{uv}\in V_1$ or $V_{vu}\cup V^{\prime}_{vu}\in V_1$ due to the complementary 
vertex pairs. 
We define a directed graph $D$ by $V(D)=V(G)$ and 
\[E(D)=\Big\{ (u,v) ~|~V_{vu}\cup V^{\prime}_{vu}\in V_1\Big\}\bigcup \Big\{ (v,u) ~|~V_{uv}\cup V^{\prime}_{uv}\in V_1\Big\}. \]
 Suppose there is a vertex $x$ in $D$ for which $w_{\mbox{out}}^x>r$.
 Clearly $x\in V_1$. We know $d_{V_1}(x)=2r+w_{\mbox{in}}^x $ and 
 $d_{V_2}(x)=2w_{\mbox{out}}^x+ w_{\mbox{in}}^x$. Then $d_{V_2}(x)> d_{V_1}(x)$, as by assumption
 $w_{\mbox{out}}^x >r$, a contradiction to the fact that $(V_1,V_2)$ is 
 a satisfactory partition of $G^{\prime}$. Hence $w_{\mbox{out}}^x\leq r$ for all
 $x\in V(D)$. \\
 
 \noindent Next we prove the following result which eliminates complementary pairs.
 
 \begin{lemma}
 {\sc  Satisfactory Partition$^{\mbox{FS}}$}, parameterized by the treewidth of the graph, is W[1]-hard.
 \end{lemma}
 
\begin{figure}[ht]
     \centering
    \[\begin{tikzpicture}[scale=1.3]
	%% Notice in the first vertex is named (v) for the sake of a later edge,
	%% and it also has a label to its left that is the math-mode $v$. 
	\vertex (v) at (1,1.5) [label=above:$\triangle^{ab}$] {};  
	\vertex (w) at (0,1) [label=left:$a$] {};
	\vertex (x) at (2,1) [label=right:$b$] {};
	\vertex (y) at (1,0.5) [label=below:$\square^{ab}$] {};
	\path
	   % Note that the word "path" here isn't used in the graph-theory sense; the \path command
	   % is always used prior to the list of edges; here, coincidentally, they do form an actual path.
		(v) edge (w)
		(w) edge (y)
		(y) edge (x)
		(v) edge (x)
	 ;   % This semicolon ends the \path command.
\end{tikzpicture}\]
\caption{Gadget for a pair of complementary vertices $(a,b)$ in the reduction from {\sc  Satisfactory Partition$^{\mbox{FSC}}$} to
{\sc  Satisfactory Partition$^{\mbox{FS}}$}.}
     \label{satfig2}
 \end{figure}
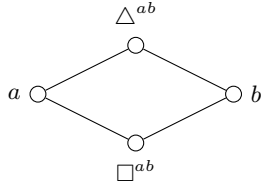

 \proof Let $I=(G,V_{\square}, V_{\triangle}, C)$ be an instance of {\sc  Satisfactory Partition$^{\mbox{FSC}}$}.
 Consider the primal graph of $I$, that is the graph $G^p$ where $V(G^p)=V(G)$ and $E(G^p)=E(G)\cup C$.
 From this  we construct an instance $I^{\prime}=(G^{\prime},V^{\prime}_{\square}, V^{\prime}_{\triangle})$ of {\sc  Satisfactory Partition$^{\mbox{FS}}$} problem. 
 For each $(a,b)\in C$ in the primal graph $G^p$, we introduce two new vertices $\triangle^{ab}$ and
 $\square^{ab}$ and  four new edges in $G^{\prime}$. We now define the $G^{\prime}$ with
 \[V(G^{\prime})=V(G) \bigcup_{(a,b)\in C} \{\triangle^{ab}, \square^{ab}\}\] and 
 \[E(G^{\prime})=E(G) \bigcup_{(a,b)\in C} \Big\{(a,\triangle^{ab}), (a, \square^{ab}), 
 (b,\triangle^{ab}), (b,\square^{ab}) \Big\}.\]
 Finally, we define the sets $V^{\prime}_{\triangle}=V_{\triangle} \bigcup_{(a,b)\in C}\{\triangle^{ab}\}$
 and $V^{\prime}_{\square}=V_{\square} \bigcup_{(a,b)\in C}\{\square^{ab}\}$. We illustrate our 
 construction in Figure \ref{satfig2}.
 It is easy to see that we can compute $I^{\prime}$ in polynomial time and 
 its treewidth is linear in the treewidth of $I$. 
 \par The following holds for every solution $(V^{\prime}_1,V^{\prime}_2)$ of $I^{\prime}$: $V^{\prime}_1$ contains
 $\triangle^{ab}$ for every $(a,b)\in C$, so it must also contain $a$ or $b$. It cannot contain both $a$ and $b$ for 
 any $(a,b)\in C$, because $\square^{ab}\in V^{\prime}_2$. Restricting  $(V^{\prime}_1, V^{\prime}_2)$  to the original
 vertices thus is a solution to $I$. Conversely, for every solution $(V_1,V_2)$ of $I$, the partition 
 $(V^{\prime}_1,V^{\prime}_2)$ where $V_1^{\prime}=V_1\bigcup_{(a,b)\in C}\{\triangle^{ab}\}$ and 
 $V_2^{\prime}=V_2\bigcup_{(a,b)\in C}\{\square^{ab}\}$, is a solution of $I^{\prime}$.
 
 This proves Theorem \ref{twtheorem}.

\section{Conclusion} In this work we proved that the {\sc Satisfactory Partition}  and   {\sc Balanced Satisfactory Partition}  problems are FPT when parameterized 
by neighbourhood diversity;  the problems are polynomial time solvable for graphs of bounded clique width, a generalized version of the {\sc Satisfactory Partition} problem  is W[1]-hard when parameterized by  treewidth.   The parameterized complexity of the {\sc Satisfactory Partition} problem 
remains unsettle  when parameterized by other important 
structural graph parameters like clique-width, modular width and treedepth.

\bibliographystyle{abbrv}
\bibliography{bibliography}
%\newpage

%\appendix

%\input{appendix}

\end{document}